\documentclass[11pt]{amsart}

\usepackage[ibidtracker=false,uniquename=false,giveninits=true,terseinits=true,backend=biber]{biblatex}
\usepackage{float}
\usepackage{graphicx}
\usepackage{todonotes}
\usepackage{subcaption}
\usepackage{amsmath}
\usepackage{amsthm}
\usepackage{amssymb}
\usepackage{algorithm}
\usepackage[noend]{algorithmic}
\usepackage[foot]{amsaddr}
\usepackage[misc]{ifsym}
\usepackage{enumitem}
\usepackage{geometry}
\usepackage[hidelinks]{hyperref}

\renewbibmacro{in:}{}
\AtEveryBibitem{
  \clearlist{language}
}

\setlist{leftmargin = 0pt}
\newtheorem{proposition}{Proposition}
\newtheorem{theorem}{Theorem}

\newtheorem{corollary}{Corollary}
\newtheorem{problem}{Problem}

\newcommand{\rnni}{\mathrm{RNNI}}
\newcommand{\findpath}{\textsc{FindPath}}

\newcommand{\rank}{\mathrm{rank}}
\newcommand{\nni}{\mathrm{NNI}}
\newcommand{\spr}{\mathrm{SPR}}
\newcommand{\tbr}{\mathrm{TBR}}
\newcommand{\fp}{\mathrm{FP}}
\newcommand{\np}{\mathbf{NP}}
\newcommand{\decprob}[1]{\rnni(#1)\text{-}\mathrm{SP}}
\renewcommand{\O}{\mathcal O}
\renewcommand{\epsilon}{\varepsilon}

\graphicspath{{figures/}}

\title[Computing $\rnni$ distance]{Computing nearest neighbour interchange distances between ranked phylogenetic trees}
\date{\today}
\author{Lena Collienne}
\email{lena.collienne@postgrad.otago.ac.nz}
\address{Department of Computer Science, University of Otago, New Zealand}
\author{Alex Gavryushkin\textsuperscript{\Letter}}
\email{\textsuperscript{\Letter}alex@biods.org}
\thanks{We thank Alexei Drummond, David Bryant, and Kieran Elmes for useful discussions about the weight difference between $\rnni$ moves, complexity, scalability, and applied aspects of our results.
Their comments improved our paper.}
\thanks{We acknowledge support from the Royal Society Te Ap\=arangi through a Rutherford Discovery Fellowship (RDF-UOO1702) awarded to AG.
This work was partially supported by Ministry of Business, Innovation, and Employment of New Zealand through an Endeavour Smart Ideas grant (CONT-61378-ENDSI-UOO) and a Data Science Programmes grant (UOAX1932).}

\begin{document}

\begin{abstract}
Many popular algorithms for searching the space of leaf-labelled (phylogenetic) trees are based on tree rearrangement operations.
Under any such operation, the problem is reduced to searching a graph where vertices are trees and (undirected) edges are given by pairs of trees connected by one rearrangement operation (sometimes called a move).
Most popular are the classical nearest neighbour interchange, subtree prune and regraft, and tree bisection and reconnection moves.
The problem of computing distances, however, is $\np$-hard in each of these graphs, making tree inference and comparison algorithms challenging to design in practice.

Although ranked phylogenetic trees are one of the central objects of interest in applications such as cancer research, immunology, and epidemiology, the computational complexity of the shortest path problem for these trees remained unsolved for decades.
In this paper, we settle this problem for the ranked nearest neighbour interchange operation by establishing that the complexity depends on the weight difference between the two types of tree rearrangements (rank moves and edge moves), and varies from quadratic, which is the lowest possible complexity for this problem, to $\np$-hard, which is the highest.
In particular, our result provides the first example of a phylogenetic tree rearrangement operation for which shortest paths, and hence the distance, can be computed efficiently.
Specifically, our algorithm scales to trees with thousands of leaves (and likely hundreds of thousands if implemented efficiently).

We also connect the problem of computing distances in our graph of ranked trees with the well-known version of this problem on unranked trees by introducing a parameter for the weight difference between move types.
We propose to study a family of shortest path problems indexed by this parameter with computational complexity varying from quadratic to $\np$-hard.
\end{abstract}

\maketitle

The problem of reconstructing evolutionary histories from sequence data is central for many popular methods in computational biology.
Most commonly trees are inferred from sequences via maximum likelihood \autocite{Stamatakis2006-xb, Guindon2010-lo}, MCMC \autocite{Ronquist2003-eq, Suchard2018-tw, Bouckaert2019-yr}, distance-, or parsimony-based approaches \autocite{Tamura2011-ky}.
All these methods rely on various tree rearrangement operations \autocite{Semple2003-nj}, the most popular of which are nearest neighbour interchange ($\nni$), subtree prune and regraft ($\spr$), and tree bisection and reconnection ($\tbr$).
Under any such operation, the tree inference problem can be formulated as a graph search, where vertices are trees and edges are given by tree rearrangement operations.
For search algorithms to be efficient, it is important to understand the geometry of these graphs.
For example, basic geometric properties of the $\nni$ graph have been successfully leveraged to speed up the maximum likelihood method \autocite{Nguyen2015-sp}.
The most basic geometric characteristic that frequently arises in applications is the minimum number of rearrangements necessary to transform one tree into another \autocite{Semple2003-nj}.
The problem then amounts to computing the length of a shortest path between trees in the three graphs.
This can also be seen as computing the distance between trees in the corresponding metric space.

Classical results in mathematical phylogenetics imply that these distances are $\np$-hard to compute for all three rearrangement operations $\nni$, $\spr$, and $\tbr$ \autocite{Dasgupta2000-xa, Bordewich2005-nx, Hickey2008-wv, Allen2001-ky}.
Intuitively, the difference between the three operations is how much change can be done to a tree by a single operation, with $\nni$ being the most local type of rearrangement and $\tbr$ the most global one.
Remarkably, it took over 25 years and a number of published erroneous attempts, as discussed in detail by \textcite{Dasgupta2000-xa}, to prove that computing distances is $\np$-hard in $\nni$ \autocite{Dasgupta2000-xa}.
Similarly, incorrect proofs for $\spr$ have been discussed in the literature \autocite{Hein1996-em, Allen2001-ky}, before \textcite{Bordewich2005-nx} proved the $\np$-hardness result for rooted trees and \textcite{Hickey2008-wv} utilised this proof to establish the result for unrooted trees.
To facilitate practical applications, fixed parameter tractable algorithms \autocite{Downey2013-nd} for computing the $\spr$ distance have been developed over the years \autocite{Whidden2010-bw, Bordewich2005-nx, Whidden2018-fw}.
Computing the $\nni$ distance is also known to be fixed parameter tractable \autocite{DasGupta1999-xf}.
Although important, these algorithms remain impractical for large distances and are only applied to trees with a moderate number of leaves or those with small distances \autocite{Whidden2018-fw}.

Another area where algorithms for computing shortest paths and distances between trees play a central role, is calculating consensus or summary trees \autocite{McMorris1994-no, Bansal2010-vr, Whidden2014-sx}.
A popular tree distance measure used in such methods is the Robinson-Foulds distance \autocite{Robinson1981-fb}, as it can be computed in linear time.
Lack of biological motivation however is a downside of this approach, which often results in poor summaries and hence is not used for summarising samples of trees obtained in a full Bayesian tree inference approach \autocite{Gavryushkin2016-uu}.
In general, distance measures that are easy to compute typically have this problem, whereas measures that are biologically relevant, including rearrangement-based distances, are often hard to compute \autocite{Whidden2018-fw}.

In this paper, we establish that using a generalisation of the $\nni$ operation introduced by \textcite{Gavryushkin2018-ol} called $\rnni$ (for Ranked Nearest Neighbour Interchange), the shortest path problem is computable in $\O(n^2)$, where $n$ is the number of tree leaves.
This makes $\rnni$ the first tree rearrangement operation under which shortest paths and distances between trees are polynomial-time computable.
Our proof of this result (Theorem~\ref{thm:rnni_polynomial}) is constructive -- we provide an algorithm called $\findpath$ that computes shortest paths in the $\rnni$ graph in $\O(n^2)$ time.
Our algorithm is optimal as shortest paths often have length quadratic in the number of leaves $n$.
The algorithm is practical as it takes seconds on a laptop to compute the distance between trees with thousands of leaves, while in the closely related $\nni$ graph the tractable number of leaves is well below twenty \autocite{Li1996-zw, Whidden2016-kl}.

Because $\nni$ can be seen as a special case of $\rnni$, we investigate whether there exists a threshold at which the complexity shifts from $\np$-hard to polynomial.
Specifically, we introduce an edge weight parameter $\rho$ in the $\rnni$ graph and consider a parametrised graph $\rnni(\rho)$.
We show that the shortest path problem is $\np$-hard in $\rnni(0)$ and quadratic in $\rnni(1)$, so the complexity changes with $\rho$.
We hence propose to characterise the complexity classes of the problem $\rnni(\rho)$ for values of $\rho \geq 0$.
This problem is similar in spirit to the beyond worst-case analysis (including parametrised complexity \autocite{Downey2013-nd}) framework \autocite{Roughgarden2019-to}.
Just like in our result, a $\gamma$-perturbation stable instance of the maximum cut problem is known \autocite{Roughgarden2019-to} to be $\np$-hard for small values of $\gamma$ and polynomial for larger values of $\gamma$.
Since the problem of identifying the value of $\gamma$ where the complexity switches has largely been resolved \autocite{Makarychev2014-ev}, we hope that the approaches reviewed by \textcite{Roughgarden2019-to} will be helpful for our proposed study as well.

\section{Definitions and background results}

\begin{figure}[ht]
\centering
\includegraphics[width=0.8\textwidth]{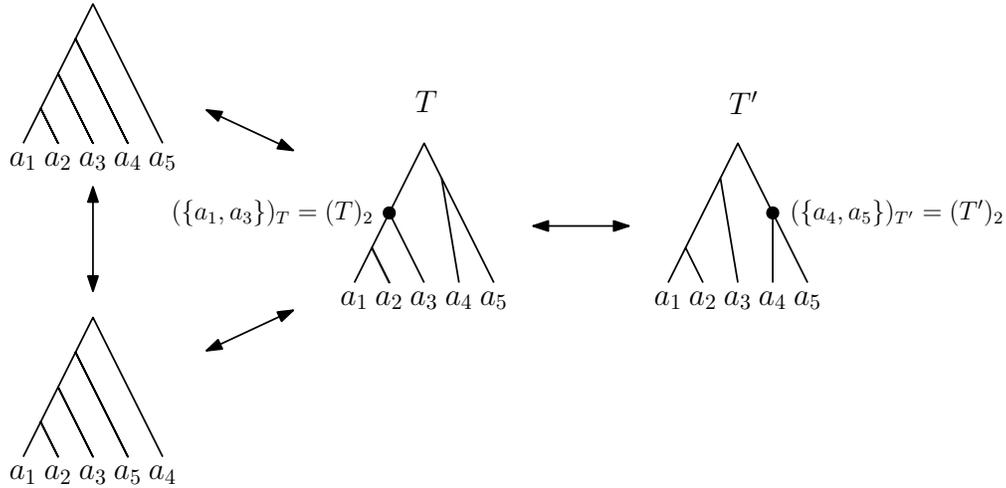}
\caption{Trees in the $\rnni$ graph with three $\nni$ moves on the left and a rank move on the right.}
\label{fig:ranked_trees_ex}
\end{figure}

Unless stated otherwise, by a \emph{tree} in this paper we mean a \emph{ranked phylogenetic tree}, which is a binary tree where leaves are uniquely labelled by elements of the set $\{a_1, \ldots, a_n\}$ for a fixed integer $n$, and all internal (non-leaf) nodes are uniquely \emph{ranked} by elements of the set $\{1, \ldots, n-1\}$ so that each child has a strictly smaller rank than its parent.
All leaves are assumed to have rank $0$ but we only refer to the ranks of internal nodes throughout.
In total there are $\frac{(n - 1)! n!}{2^{n-1}}$ such trees on $n$ leaves \autocite{Gavryushkin2018-ol}.
Two trees are considered to be identical if there exists an isomorphism between them which preserves edges, leaf labels, and node rankings.
For example, trees in Figure~\ref{fig:ranked_trees_ex} are all different.

Because internal nodes of a tree $T$ are ranked uniquely, we can address \textbf{the} node of rank ${t \in \{1, \ldots, n - 1\}}$, and we write $(T)_t$ to denote this node.
An \emph{interval} $[(T)_t,(T)_{t+1}]$ is defined by two nodes of consecutive ranks.
A \emph{cluster} $C \subseteq \{a_1, \ldots, a_n\}$ in a tree $T$ is a subset of leaves that contains all leaves descending from one internal node of $T$.
We then say that this internal node \emph{induces} the cluster $C$, and that the subtree rooted at this node is \emph{induced} by $C$.
Trees can uniquely be specified using the \emph{cluster representation}, that is a list of all clusters induced by internal nodes of that tree ordered according to the ranks of internal nodes.
For example, the cluster representation of tree $T$ in Figure~\ref{fig:ranked_trees_ex} is $[\{a_1, a_2\}, \{a_1, a_2, a_3\}, \{a_4, a_5\}, \{a_1,a_2,a_3,a_4,a_5\}]$.
For a set $S \subseteq \{a_1, \ldots, a_n\}$ and tree $T$ we denote the \emph{most recent common ancestor} of $S$ in $T$, that is the node of the lowest rank in $T$ that induces a cluster containing all elements of $S$, by $(S)_T$.
Note that $(C)_T = (T)_t$ if the cluster $C$ is induced by the node of rank $t$ in $T$.

Our main object of study is the following class of graphs $\rnni(\rho)$ indexed by a real-valued parameter $\rho \geq 0$.
Vertices of the $\rnni(\rho)$ graph are trees as defined above.
Two trees are connected by an edge (also called an \emph{$\rnni$ move}) if one results from the other by performing one of the following two types of tree rearrangement operation (see Figure~\ref{fig:ranked_trees_ex}):\\
(i) A \emph{rank move} on a tree $T$ exchanges the ranks of two internal nodes $(T)_t$ and $(T)_{t+1}$ with consecutive ranks, provided the two nodes are not connected by an edge in $T$.\\
(ii) Trees $T$ and $R$ are connected by an \emph{$\nni$ move} if there are edges $e$ in $T$ and $f$ in $R$ both connecting nodes of consecutive ranks in the corresponding trees, such that the (non-binary) trees obtained by shrinking $e$ and $f$ into internal nodes are identical.\\
The parameter $\rho \geq 0$ is the weight of the rank move operation, an $\nni$ move weighs $1$.

The \emph{weight of a path} in $\rnni(\rho)$ is the sum of the weights of all moves along the path.
The \emph{distance} between two trees in $\rnni(\rho)$ is the weight of a path with the minimal weight, which we will call a \emph{shortest path}.
When $\rho = 1$ we assume that the graph is unweighted.

We consider the following class of problems parametrised by a real number $\rho \geq 0$.

\noindent\fbox{\parbox{\textwidth}{
$\decprob{\rho}$\\
INSTANCE: A pair of trees $T$ and $R$ on $n$ leaves\\
FIND: A path of minimal weight between $T$ and $R$ in $\rnni(\rho)$
}}

Since $\rnni(\rho)$ is a connected graph, there always exists a solution to $\decprob{\rho}$.
Furthermore, the size of every solution to an instance of $\decprob{\rho}$ is bounded by a polynomial in $n$, despite the search space being super-exponential.
This is because the diameter of the $\rnni(1)$ graph is bounded from above \autocite{Gavryushkin2018-ol} by $n^2 - 3n - 5/8$.

Our main goal is to prove that $\decprob{1}$ can be solved in polynomial time.
We will see later in the paper that it follows from a classical result \autocite{Dasgupta2000-xa} that $\decprob{0}$ is $\np$-hard.
To be consistent with notations used in the literature \autocite{Gavryushkin2018-ol}, we will denote the graph $\rnni(1)$ by $\rnni$.

\section{$\findpath$ algorithm}
\label{sec:rnni_complexity}

In this section we introduce an algorithm called $\findpath$ that computes paths between trees and is quadratic in the number of leaves.

An input of the $\findpath$ algorithm is two trees $T$ and $R$ in their cluster representation.
We denote the representation of $R$ by $[C_1, \ldots, C_{n-1}]$.
The algorithm considers the clusters $C_1, \ldots, C_{n-2}$ iteratively in their order and produces a sequence $p$ of trees which becomes a shortest path from $T$ to $R$ after the algorithm terminates.
During each iteration $k = 1, \ldots, n-2$ new trees are added to $p$ if necessary, and we will refer to the last added tree as $T_1$.
In iteration $k$, the rank of $(C_{k})_{T_1}$ is decreased by $\rnni$ moves until $C_k$ is induced by the node of rank $k$ in $T_1$.
In Proposition~\ref{prop:fp_correctness} we show that $\findpath$ is a deterministic algorithm with running time quadratic in the number of leaves $n$.
In particular, there always exists a unique move that decreases the rank of $(C_{k})_{T_1}$ as described above.

\begin{algorithm}[H]
\caption{$\findpath$($T,R$)}
\begin{algorithmic}[1]
\STATE $T_1 := T$, $p := [T_1]$, $[C_1, \ldots, C_{n-1}] := R$
\FOR {$k = 1, \dots, n-2$}
\label{alg:findpath:line:for_loop}
	\WHILE {$\rank((C_k)_{T_1})>k$}
	\label{alg:findpath:line:while_loop}
		\IF {$(C_k)_{T_1}$ and node $u$ preceding $(C_k)_{T_1}$ in $T_1$ are connected by an edge}
			\STATE $T_2$ is $T_1$ with the rank of $(C_k)_{T_1}$ decreased by an $\nni$ move
			\label{alg:nni_step}
		\ELSE
			\STATE $T_2$ is $T_1$ with ranks of $u$ and $(C_k)_{T_1}$ swapped
			\label{alg:rank_step}
		\ENDIF
		\STATE $T_1 = T_2$
		\STATE $p = p+T_1$
	\ENDWHILE
\ENDFOR
\RETURN $p$
\end{algorithmic}
\end{algorithm}

\begin{proposition}
$\findpath$ is a correct deterministic algorithm that runs in $\O(n^2)$ time.
\label{prop:fp_correctness}
\end{proposition}

\proof
To show that $\findpath$ is a deterministic algorithm (see the pseudocode above), we have to prove that tree $T_2$ constructed in the \textbf{while} loop (line~\ref{alg:findpath:line:while_loop}) of the algorithm always exists and is uniquely defined.
If $T_2$ is obtained in line~\ref{alg:rank_step} from $T_1$ by a rank move, the tree exists and is unique because there always exists exactly one rank move on any particular interval that is not an edge.
It remains to show that an $\nni$ move that decreases the rank of $(C_k)_{T_1}$ always exists and is unique.
To prove this we consider cases $k = 1$ and $k > 1$ separately.

\begin{itemize}
\item[Case $k = 1$.]
In this case $C_k$ consists of two leaves $\{x, y\}$.
Since we assumed that the \textbf{while} condition is satisfied, the node $v = (\{x, y\})_{T_1}$ has rank $r > 1$.
Consider the node $u$ preceding $v$ in $T_1$, so the rank of $u$ is $r - 1$.
Assume without loss of generality that $x$ is in the cluster induced by $u$, so $y$ has to be outside this cluster.
Consider the following three disjoint subtrees of $T_1$: the subtree $T_{11}$ induced by a child of $u$ and containing $x$, the subtree $T_{12}$ induced by the other child of $u$, the subtree $T_{13}$ induced by a child of $v$ and containing $y$.
Now observe that out of two $\nni$ moves possible on the edge $[u, v]$ in $T_1$, only the one that swaps $T_{12}$ and $T_{13}$ does decrease the rank of the most recent common ancestor of $\{x, y\}$.
Hence $T_2$ exists and is unique in this case.

\item[Case $k > 1$.]
In this case $C_k = C_i \cup C_j$ for $i, j < k$.
In this case the subtree of $T_1$ induced by $(C_i)_{T_1}$ is identical to the subtree of $R$ induced by $(C_i)_R$, and the same is true for $(C_j)_{T_1}$ and $(C_j)_R$.
Hence, we can reduce this case to $k = 1$ by suppressing $C_i$ and $C_j$ in both $T_1$ and $R$ to new leaves $c_i$ and $c_j$ (of rank zero) respectively.
As in Case $k = 1$, exactly one of two possible $\nni$ moves deceases the rank of the most recent common ancestor of $c_i$, $c_j$ in $T_1$, so the same is true for the most recent common ancestor $(C_k)_{T_1}$, and $T_2$ is unambiguously defined.
\end{itemize}

Thus, $\findpath$ is a deterministic algorithm.

To prove correctness, note that the algorithm starts by adding $T$ to the output path, and every new tree added to the output path is an $\rnni$ neighbour of the previously added one (see line~\ref{alg:nni_step} and \ref{alg:rank_step}).
To see that the output path terminates in $R$, observe that after $k$ iteration of the \textbf{for} loop (line~\ref{alg:findpath:line:for_loop}) of the algorithm, the first $k$ clusters of $T_1$ and $R$ must coincide, and so after $n-2$ iterations a path between $T$ and $R$ is constructed.

The worst-case time complexity of $\findpath$ is quadratic in the number of leaves, as there can be at most $n-2$ executions of the \textbf{for} loop (line~\ref{alg:findpath:line:for_loop}) and in every iteration of the \textbf{for} loop at most $n-2$ \textbf{while} loops (line~\ref{alg:findpath:line:while_loop}) are executed.
Here and throughout the paper we assume that the output of $\findpath$ is encoded by a list of $\rnni$ moves rather than an actual list of trees.
This is because writing out a tree on $n$ leaves takes time linear in $n$ and the complexity of $\findpath$ becomes cubic.
\endproof

\section{$\findpath$ computes shortest paths in optimal time}

In this section we prove the main result of this paper, that $\decprob{1}$ is polynomial.
Specifically we prove that paths returned by $\findpath$ are always shortest.
We also show that $\findpath$ is an optimal algorithm, that is, no sub-quadratic algorithm can solve $\decprob{1}$.

The main ingredient of our proof is to show that a local property (see~(\ref{eqn:iff_inequality}) in the proof) of the $\findpath$ algorithm is enough to establish that the output paths are shortest.
The property can intuitively be understood as $\findpath$ always choosing the best tree possible to go to.
Importantly, this result can be used for an arbitrary vertex proposal algorithm in an arbitrary graph to establish that the algorithm always follows a shortest path between vertices in the graph, hence our proof technique is of general interest.

\begin{theorem}
The worst-case time complexity of the shortest path problem in the $\rnni$ graph on trees with $n$ leaves is $\O(n^2)$.
Hence $\decprob{1}$ is polynomial time solvable.
\label{thm:rnni_polynomial}
\end{theorem}

\proof
We prove this theorem by showing that for every pair of trees $T$ and $R$, the path computed by the $\findpath$ algorithm is a shortest $\rnni$ path.
We denote this path by $\fp(T, R)$ and its length by $|\fp(T, R)|$.
By $d(T, R)$ we denote the length of a shortest path between $T$ and $R$, that is, the $\rnni$ distance between trees.
We hence want to show that $|\fp(T, R)| = d(T, R)$ for all trees.

Assume to the contrary that $T$ and $R$ are two trees with a minimum distance $d(T, R)$ such that $d(T,R) \neq |\fp(T,R)|$, that is, $d(T,R) < |\fp(T,R)|$.
Let $T'$ be the first tree on a shortest $\rnni$ path from $T$ to $R$.
Then $d(T',R) = d(T, R) - 1$, implying that the distance between $T'$ and $R$ is strictly smaller than that between $T$ and $R$.
Hence $d(T', R) = |\fp(T',R)| < |\fp(T,R)| - 1$.
We finish the proof by showing that no trees satisfy this inequality.

Specifically, we will show that
\begin{equation}
\begin{split}
\mbox{for all trees $T$, $R$, and $T'$}	& \mbox{ such that $T'$ is one $\rnni$ move away from $T$,}\\
					&|\fp(T',R)| \geq |\fp(T,R)| - 1
\end{split}
 \label{eqn:iff_inequality}
\end{equation}

\begin{figure}[ht]
\centering
\includegraphics[width=0.5\textwidth]{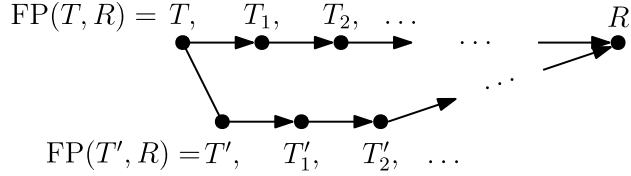}
\caption{Trees $T$, $T'$, and $R$ as in inequality~(\ref{eqn:iff_inequality}).
Paths $\fp(T,R) = [T,T_1,T_2, \ldots, R]$ and $\fp(T',R) = [T',T'_1,T'_2, \ldots, R]$ are indicated by arrows.}
\label{fig:proof_idea}
\end{figure}

We will use Figure~\ref{fig:proof_idea} to demonstrate our argument.

Assume to the contrary that $T$ and $R$ are trees for which there exists $T'$ violating inequality~(\ref{eqn:iff_inequality}).
Out of all such pairs $T, R$ choose one with the minimal $|\fp(T, R)|$.
Denote $\fp(T,R) = [T, T_1, T_2, \ldots, R]$ and $\fp(T', R) = [T', T'_1, T'_2, \ldots, R]$, and let $[(T)_t, (T)_{t+1}]$ be the interval in $T$ on which the $\rnni$ move connecting $T$ and $T'$ is performed.
Let $C_k$ be the cluster of $R$ such that the node $(C_k)_T$ is moved down by the first move on $\fp(T, R)$.
If the rank of $(C_k)_T$ is not in $\{t, t+1\}$ then $(C_k)_T$ and $(C_k)_{T'}$ induce the same cluster, so $\findpath$ would make the same rearrangement in both trees $T$ and $T'$ in the first move along $\fp(T, R)$ and $\fp(T', R)$ resulting in trees $T_1$ and $T'_1$ which are $\rnni$ neighbours, as in Figure~\ref{fig:proof_idea}.
In this case, paths $\fp(T_1, R)$ and $\fp(T'_1, R)$ violate inequality~(\ref{eqn:iff_inequality}) but $\fp(T_1, R)$ is strictly shorter than $\fp(T, R)$, contradicting our minimality assumption.
Hence, the first move on $\fp(T, R)$ has to involve an interval incident to at least one of the nodes $(T)_t$, $(T)_{t+1}$.

Moreover, because $C_k$ is the first cluster satisfying the \textbf{while} condition of $\findpath$ applied to $T$ and $R$, all clusters $C_j$ with $j < k$ have to be present in $T$.
And since the first move on $\fp(T,R)$, which decreases the rank of $(C_k)_T$, involves nodes with ranks not higher than $t+2$, the most recent common ancestor of $C_k$ has rank not higher than $t+1$ after this move.
Hence $k \leq t + 1$.
Furthermore, clusters $C_j$ for all $j \leq k - 2$ have to be present in $T'$ as well as $T$, because all clusters induced by nodes of rank $t - 1$ or lower coincide in these two trees.
Cluster $C_{k-1}$, however, might not be induced by a node in $T'$ if $k-1 = t$.
Therefore, the first move on $\fp(T', R)$ can decrease the rank of the most recent common ancestor of either $C_{k-1}$ or $C_k$.

We will distinguish two cases depending on whether $T$ and $T'$ are connected by an $\nni$ or a rank move.
For each of these we will further distinguish all possible moves between $T$ and $T_1$.
Note that in all figures illustrating possible moves on $\fp(T,R)$ and $\fp(T',R)$ below, the position of the tree root is irrelevant, so we have positioned roots to simplify our figures.

\begin{figure}[ht]
\centering
\includegraphics[width=0.35\textwidth]{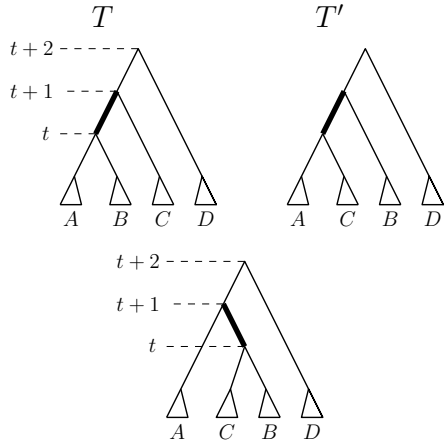}
\caption{$\nni$ move between $T$ and $T'$ on the edge $[(T)_t,(T)_{t+1}]$ indicated in bold, and the third $\rnni$ neighbour resulting from a move on this edge.}
\label{fig:thm_fp_nni1}
\end{figure}

\begin{enumerate}
\item[\textbf{Case 1.}]
$T$ and $T'$ are connected by an $\nni$ move.
So $[(T)_t,(T)_{t+1}]$ is an edge in $T$ -- see Figure~\ref{fig:thm_fp_nni1}.
Denote the clusters induced by the children of $(T)_t$ by $A$ and $B$ and the cluster induced by the child of $(T)_{t+1}$ that is not $(T)_t$ by $C$, and assume that the $\nni$ move between $T$ and $T'$ exchanges the subtrees induced by clusters $B$ and $C$.
Additionally, denote the cluster induced by the child of $(T)_{t+2}$ that is not $(T)_{t+1}$ by $D$ -- see Figure~\ref{fig:thm_fp_nni1}.
Note that $[(T)_{t+1}, (T)_{t+2}]$ does not have to be an edge in tree $T$ (see Case~\ref{case:rank_move_interval_above}).
\end{enumerate}

We now consider all possible moves $\findpath$ can perform to go from $T$ to $T_1$ that involve a node of rank $t$ or $t+1$, that is, we will consider three intervals in total.

\begin{enumerate}[label = 1.{\arabic*}]
\item $\rnni$ move (either type) on interval $[(T)_t, (T)_{t+1}]$.
This move has to be the $\nni$ move that is different from the $\nni$ move connecting $T$ and $T'$.
In this case, the cluster $B \cup C$ is built in $T_1$, as depicted in the bottom of Figure~\ref{fig:thm_fp_nni1}.
Hence the first cluster $C_k$ that satisfies the \textbf{while} condition of $\findpath$ must contain elements from both $B$ and $C$ but not from $A$, and the rank of $(C_k)_R$ has to be at most $t$.
But then $\findpath$ applied to $T'$ and $R$ has to decrease the rank of $(C_k)_{T'}$ in its first step implying that $T'_1 = T_1$, so $|\fp(T',R)| = |\fp(T,R)|$.
This contradicts our assumption that $|\fp(T',R)| < |\fp(T,R)| - 1$.

\item $\nni$ move on (edge) interval $[(T)_{t+1}, (T)_{t+2}]$ that swaps the subtrees induced by clusters $C$ and $D$.
This move is shown in Figure~\ref{fig:thm_fp_nni2a} by an arrow from $T$ to the leftmost tree in the middle row.
In this case, the first cluster $C_k$ that satisfies the \textbf{while} condition of $\findpath$ computing $\fp(T, R)$ must intersect $D$ but not $C$.
Additionally, $C_k$ must intersect $A$, or $B$, or both of them.
Hence, we will consider each of these three cases individually, and demonstrate them in Figure~\ref{fig:thm_fp_nni}.

\begin{figure}[ht]
	\begin{subfigure}[b]{.45\textwidth}
		\centering
		\includegraphics[width=0.9\linewidth]{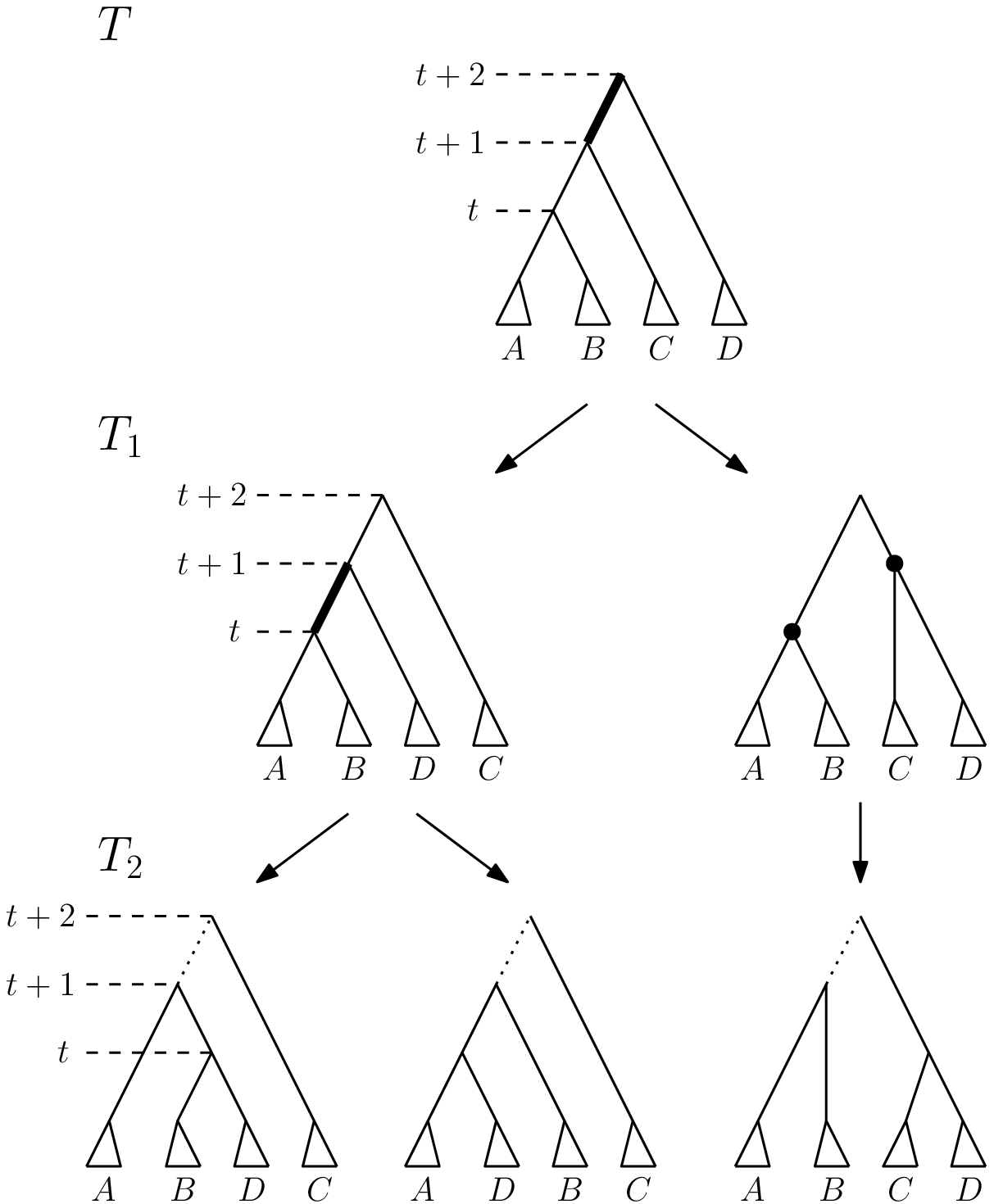}
		\vspace{12pt}
		\caption{Possible initial segments of $\fp(T, R)$}
		\label{fig:thm_fp_nni2a}
	\end{subfigure}
	\begin{subfigure}[b]{.45\textwidth}
		\centering
		\includegraphics[width=0.9\linewidth]{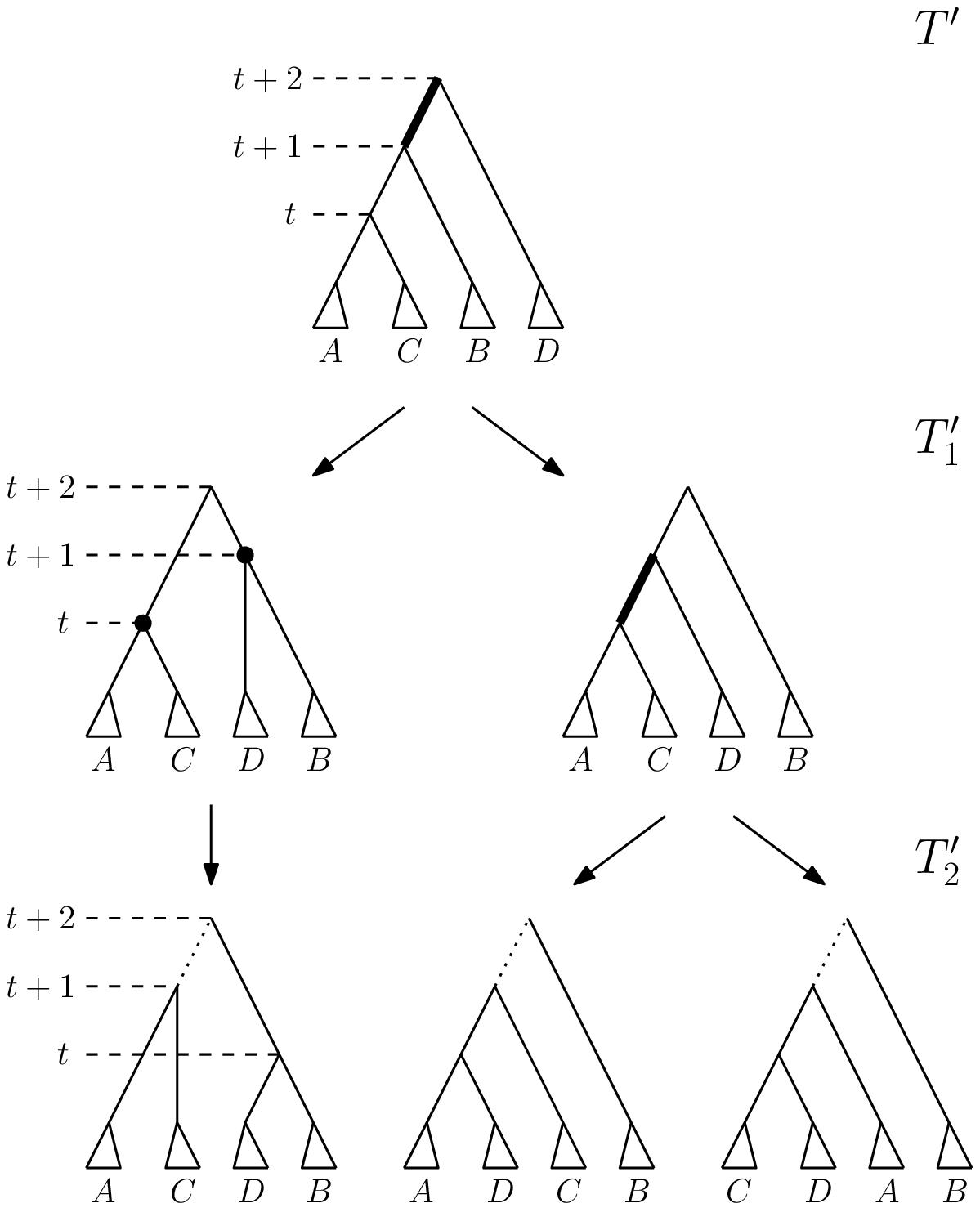}
		\vspace{12pt}
		\caption{Possible initial segments of $\fp(T', R)$}
		\label{fig:thm_fp_nni2b}
	\end{subfigure}
	\caption{Comparison of paths $\fp(T, R)$ and $\fp(T', R)$ if $T$ and $T'$ are connected by an $\nni$ move on edge $[(T)_{t},T_{t+1}]$ in $T$.
	The bottom row displays all possibilities for $T_2$ and $T'_2$, depending on the position of cluster $C_k$ that satisfies the \textbf{while} condition of $\findpath$:
	case $C_k$ intersects $B$ and $D$ is on the left, $C_k$ intersects $A$ and $D$ is in the middle, and $C_k$ intersects $C$ and $ D$ is on the right.}
	\label{fig:thm_fp_nni}
\end{figure}

\begin{enumerate}[label = \theenumi.\arabic*]
\item $C_k$ intersects $A$, $B$, and $D$ but not $C$.
In this case, since we assumed $[(T_1)_t, (T_1)_{t+1}]$ to be an edge in the tree, no move on $T_1$ can decrease the rank of $(C_k)_{T_1}$.
It follows from the proof of Proposition~\ref{prop:fp_correctness} that this can happen only when the subtrees induced by $(C_k)_{T_1}$ and $(C_k)_R$ in the corresponding trees coincide.
That is, the \textbf{while} condition of $\findpath$ must be false after this first move for all $j \leq k$.
This implies that $t = k - 1$ and $C_{k-1} = A \cup B$.
But since the rank of $(C_{k-1})_{T'}$ is $t + 1 > k - 1$, $C_{k-1}$ has to be the first cluster for which the \textbf{while} condition of $\findpath$ applied to $T'$ and $R$ is met.
Hence the first move on $\fp(T', R)$ must decrease the rank of $(C_{k-1})_{T'}$ by building the cluster $A \cup B$, in which case $T'_1 = T$.
This however contradicts $|\fp(T',R)| < |\fp(T,R)| - 1$.

\item $C_k$ intersects $A$ and $D$ but not $B$ or $C$.
\label{deep_case_details}
Starting from $T$, $\findpath$ exchanges first subtrees induced by clusters $C$ and $D$ and then by $B$ and $D$.
This results in trees $T_1$ and $T_2$ -- see the path leading to the tree in the middle of the bottom row in Figure~\ref{fig:thm_fp_nni2a}.
This implies that the rank of $(C_{k-1})_R$ is lower than $t$, so the first cluster that satisfies the \textbf{while} condition of $\findpath$ applied to $T'$ and $R$ is $C_k$.
Hence, starting from $T'$, $\findpath$ exchanges first subtrees induced by $B$ and $D$ and then by $C$ and $D$.
This results in trees $T'_1$ and $T'_2$ -- see the path leading to the tree in the middle of the bottom row in Figure~\ref{fig:thm_fp_nni2b}.
It follows that $T_2$ and $T'_2$ are connected by an $\rnni$ move on the interval $[(T_2)_{t+1}, (T_2)_{t+2}]$ (indicated by dotted edges in the corresponding trees in Figure~\ref{fig:thm_fp_nni}).
This together with the facts that $|\fp(T_2, R)| = |\fp(T, R)| - 2$ and $|\fp(T'_2, R)| = |\fp(T', R)| - 2$ contradicts the assumption that $\fp(T, R)$ is of minimal length violating inequality~(\ref{eqn:iff_inequality}).

\item $C_k$ intersects $B$ and $D$ but not $A$ or $C$.
This case is analogous to the previous one.
The two initial segments of $\fp(T, R)$ and $\fp(T', R)$ are the paths leading to the leftmost trees in the bottom row of Figures~\ref{fig:thm_fp_nni2a} and \ref{fig:thm_fp_nni2b}, respectively.
Note that the rank swap leading from $T'_1$ to $T'_2$ is required because the rank of $(C_k)_R$ is at most $t$ as implied by the move leading from $T_1$ to $T_2$.
The corresponding trees $T_2$ and $T'_2$ are again $\rnni$ neighbours.
\end{enumerate}

\item $\nni$ move on (edge) interval $[(T)_{t+1}, (T)_{t+2}]$ that builds a cluster $C \cup D$ in $T_1$.
\label{case:one_or_two_moves_down}
This move is shown in Figure~\ref{fig:thm_fp_nni2a} by an arrow from $T$ to the second leftmost tree in the middle row.
In this case, $C_k$ intersects $C$ and $D$ but not $A$ or $B$.
And we have the following two possibilities to consider.

\begin{enumerate}[label = \theenumi.\arabic*]
\item The ranks of $(C_k)_{T_1}$ and $(C_k)_R$ coincide.
In this case, the previous cluster $C_{k-1}$ of $R$ has to be $A \cup B$.
Since $A \cup B$ is not a cluster in $T'$, the first $\rnni$ move on $\fp(T', R)$ builds the cluster $A \cup B$ by swapping subtrees induced by cluster $B$ and $C$.
This move results in $T'_1 = T$ contradicting $|\fp(T',R)| < |\fp(T,R)| - 1$.

\begin{figure}[ht]
	\centering
	\includegraphics[width=0.4\textwidth]{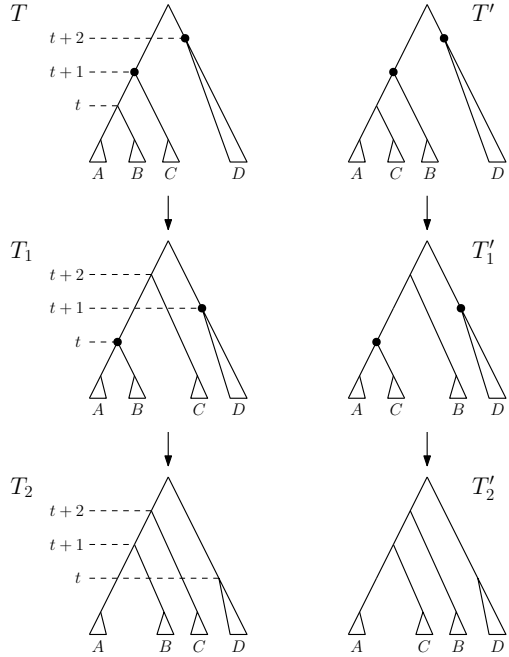}
	\caption{Comparison of paths $\fp(T, R)$ and $\fp(T', R)$ if there is an $\nni$ move between $T$ and $T'$ and a rank move on the interval above this edge follows on $\fp(T, R)$.}
	\label{fig:thm_fp_nni3}
\end{figure}

\item The rank of $(C_k)_{T_1}$ is strictly higher than that of $(C_k)_R$.
In this case, $\findpath$ decreases the rank of $(C_k)_{T_1}$ in the second step.
This results in the path from $T$ to the rightmost tree in Figure~\ref{fig:thm_fp_nni2a}.
Hence, $\fp(T', R)$ also has to begin with two moves that decrease the rank of $(C_k)_{T'}$ twice, resulting in the rightmost path in Figure~\ref{fig:thm_fp_nni2b}.
Similarly to case~\ref{deep_case_details}, we arrive at a contradiction that trees $T_2$, $T'_2$, and $R$ violate inequality~(\ref{eqn:iff_inequality}) and $|\fp(T_2,R)| < |\fp(T,R)|$.
\end{enumerate}

\item Rank move on interval $[(T)_{t+1},(T)_{t+2}]$.
\label{case:rank_move_interval_above}
This case is analogous to case~\ref{case:one_or_two_moves_down} (see Figure~\ref{fig:thm_fp_nni3}).
If the ranks of $(C_k)_{T_1}$ and $(C_k)_R$ coincide then $C_{k-1} = A \cup B$, and applying $\findpath$ to $T', R$ we get $T'_1 = T$.
If the rank of $(C_k)_{T_1}$ is strictly higher than that of $(C_k)_R$ then $\findpath$ decreases the rank of $(C_k)_{T_1}$ in the second step.
Recall that the interval between nodes of rank $t$ and $t+1$ is an edge in both $T$ and $T'$.
Hence, the first two moves on $\fp(T', R)$ decrease the rank of $(C_k)_{T'}$ twice resulting in $T'_2$ which is an $\rnni$ neighbour of $T_2$ as depicted in Figure~\ref{fig:thm_fp_nni3}.
As before, this contradicts our minimality assumption.

\item $\rnni$ move (either type) on interval $[(T)_{t-1},(T)_t]$.
In this case $C_k \subseteq A \cup B$ and the rank of $(C_k)_{R}$ is at most $t - 1$.
This implies that $C_k$ is the first cluster to satisfy the \textbf{while} condition for $T'$ and the first move on $\fp(T', R)$ decreases the rank of $(C_k)_{T'}$ by exchanging the subtrees induced by $B$ and $C$.
This results in $T'_1 = T$.
\end{enumerate}

\begin{figure}[ht]
\centering
\includegraphics[width=0.4\textwidth]{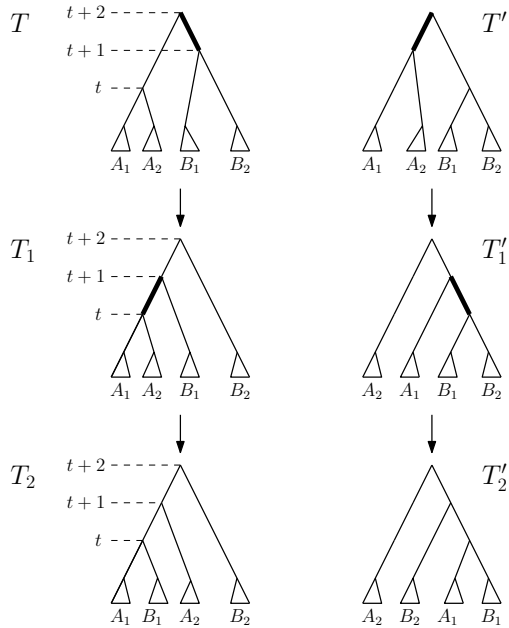}
\caption{Rank move between $T$ and $T'$ and possible initial segments of $\fp(T, R)$ and $\fp(T', R)$ when $[(T)_{t+1}, (T)_{t+2}]$ is an edge.
We use notations ${A = A_1 \cup A_2}$ and $B = B_1 \cup B_2$.}
\label{fig:thm_fp_rank1}
\end{figure}

\begin{enumerate}
\item[\textbf{Case 2.}] $T$ and $T'$ are connected by a rank move.
We assume that the rank move is performed on the interval $[(T)_t, (T)_{t+1}]$.
Denote the cluster induced by $(T)_t$ by $A$, the clusters induced by the children of $(T)_t$ by $A_1$ and $A_2$, the cluster induced by $(T)_{t+1}$ by $B$, and the clusters induced by the children of $(T)_{t+1}$ by $B_1$ and $B_2$ -- see Figure~\ref{fig:thm_fp_rank1}.
\end{enumerate}

We again consider all possible moves $\findpath$ can perform to go from $T$ to $T_1$ that involve a node of rank $t$ or $t+1$.

\begin{enumerate}[label = 2.\arabic*]
\item Rank move on $[(T)_t,(T)_{t+1}]$.
This move results in $T_1 = T'$.

\item $\nni$ move on (edge) interval $[(T)_{t+1},(T)_{t+2}]$.
The following two sub-cases are analogous to case~\ref{case:one_or_two_moves_down}.

\begin{enumerate}[label = \theenumi.\arabic*]
\item $(T)_{t+2}$ is a parent of $(T)_t$.
The first move on $\fp(T, R)$ builds a cluster $A \cup B_1$ or $A \cup B_2$, and we assume without loss of generality that it is the former, as in Figure~\ref{fig:thm_fp_rank1}.
This implies that $C_k$ intersects $A$ and $B_1$ but not $B_2$
If the ranks of $(C_k)_{T_1}$ and $(C_k)_R$ coincide then the previous cluster $C_{k-1}$ of $R$ has to be $A$.
Therefore, the first move on $\fp(T', R)$ decreases the rank of $(A)_{T'}$, which results in $T'_1 = T$.
If the rank of $(C_k)_{T_1}$ is strictly higher than that of $(C_k)_R$ then $\findpath$ decreases the rank of $(C_k)_{T_1}$ in the second step.
Due to the symmetry we can assume that $C_k \subseteq A_1 \cup B_1$, which implies that the move between $T_1$ and $T_2$ exchanges the subtrees induced by $A_2$ and $B_1$, as depicted on the left of Figure~\ref{fig:thm_fp_rank1}.
$C_k \subseteq A_1 \cup B_1$ implies that the first two moves on $\fp(T', R)$ result in a tree $T'_2$ that is an $\rnni$ neighbour of $T_2$ -- see Figure~\ref{fig:thm_fp_rank1}.
This is a contradiction to the minimality assumption on $|\fp(T,R)|$.

\item $(T)_{t+2}$ is not a parent of $(T)_t$.
In this case, there exists a cluster $C$ induced by the child of $(T)_{t+2}$ which is different from the one that induces $B$ -- see Figure~\ref{fig:thm_fp_rank2}.
\begin{figure}[ht]
\centering
\includegraphics[width=0.4\textwidth]{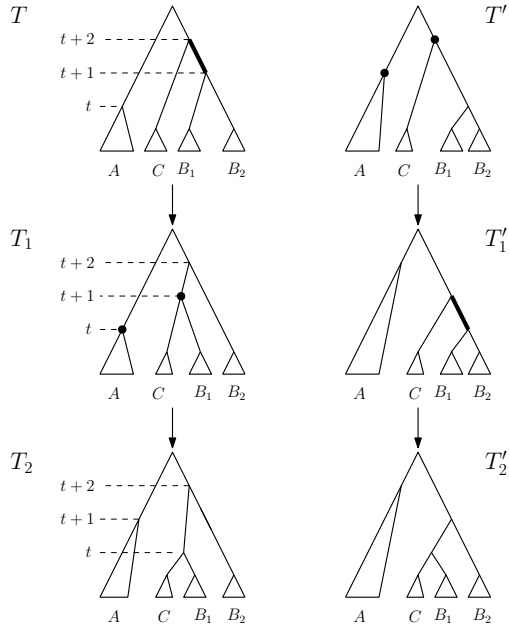}
\caption{Comparison of paths $\fp(T, R)$ and $\fp(T', R)$ if there is a rank move between $T$ and $T'$ and an $\nni$ move on the edge below the corresponding (rank) interval follows on $\fp(T, R)$.}
\label{fig:thm_fp_rank2}
\end{figure}
We can assume without loss of generality that $C_k \subseteq C \cup B_1$ and the first move on $\fp(T, R)$ builds a new cluster $C \cup B_1$.
If the ranks of $(C_k)_{T_1}$ and $(C_k)_R$ coincide then $C_{k-1} = A$, which implies that $A$ is induced by the node of rank $t$ in both $T$ and $R$.
So $T'_1 = T$.
If the rank of $(C_k)_{T_1}$ is strictly higher than that of $(C_k)_R$ then $\findpath$ decreases the rank of $(C_k)_{T_1}$ in the second step -- see Figure~\ref{fig:thm_fp_rank2}.
The corresponding first moves on $\fp(T', R)$ are shown on the right in Figure~\ref{fig:thm_fp_rank2}, and we again get that $T_2$ and $T'_2$ are $\rnni$ neighbours.
\end{enumerate}

\item Rank move on interval $[(T)_{t+1}, (T)_{t+2}]$.
Again, depending on whether or not the ranks of $(C_k)_{T_1}$ and $(C_k)_R$ coincide, we arrive at the conclusion that either $T'_1 = T$ or $T_2$ and $T'_2$ are $\rnni$ neighbours, similarly to case~\ref{case:rank_move_interval_above}.

\item $\rnni$ move (either type) on interval $[(T)_{t-1},(T)_t]$.
In this case $C_k \subseteq A$ and the first move on $\fp(T', R)$ must be a rank swap resulting in $T'_1 = T$.
\end{enumerate}

Since all possible cases result in a contradiction, we conclude that inequality~(\ref{eqn:iff_inequality}) is true for all trees, which completes the proof of the theorem.
\endproof

We finish this section by showing that no algorithm has strictly lower worst-case time complexity than $\findpath$.
We again assume here that the output of an algorithm for solving $\decprob{1}$ is a list of $\rnni$ moves.
Requiring the output to be a list of trees would result in cubic complexity while maintaining the optimality of $\findpath$.

\begin{corollary}
The time-complexity of the shortest path problem $\decprob{1}$ is $\Omega(n^2)$.
\end{corollary}

\proof
We prove this by establishing the lower bound on the output size to the problem, that is, the length of a shortest paths.

Consider two ``caterpillar" trees $T = [\{a_1, a_2\}, \{a_1, a_2, a_3\}, \ldots, \{a_1, a_2, \ldots, a_n\}]$ and $R = [\{a_1,a_n\}, \{a_1, a_n, a_{n-1}\}, \ldots, \{a_1, a_n, \ldots, a_2\}]$.
Applied to these trees $\findpath$ executes an $\nni$ move in each of the $n-k-1$ \textbf{while} loops (line~\ref{alg:findpath:line:while_loop}) in every iteration $k$ of the \textbf{for} loop (line~\ref{alg:findpath:line:for_loop}).
Hence the length of the output path of $\findpath$ is $\sum\limits_{k = 1}^{n - 2} k = \frac{(n-1)(n-2)}{2}$ and therefore quadratic in $n$.
Theorem~\ref{thm:rnni_polynomial} then implies that this path is a shortest path.
It follows that the worst-case size of the output to $\decprob{1}$ is quadratic.
\endproof

\section{For what $\rho$ is $\decprob{\rho}$ polynomial?}

As we have seen in Section~\ref{sec:rnni_complexity}, the shortest path problem $\decprob{1}$ is solvable in polynomial time.
In this section, we will show that a classical result in mathematical phylogenetics implies that $\decprob{0}$ is $\np$-hard.
We will also discuss $\decprob{\rho}$ for other values of $\rho$.

\begin{theorem}[\textcite{Dasgupta2000-xa}]
$\decprob{0}$ is $\np$-hard.
\label{thm:nni_hard}
\end{theorem}

\proof
Note that the length of the path required in an instance of $\decprob{0}$ is equal to the minimum number of $\nni$ moves necessary to convert one tree into another tree where the rankings of internal nodes are ignored and $\nni$ moves are allowed on every edge.
This minimum is called the $\nni$ distance and the corresponding problem is known to be $\np$-hard \autocite{Dasgupta2000-xa}.
\endproof

In the light of Theorem~\ref{thm:rnni_polynomial} and Theorem~\ref{thm:nni_hard} the following problem is natural.

\begin{problem}
Characterise the complexity of $\decprob{\rho}$ in terms of $\rho$.
\label{prblm:rho_range}
\end{problem}

This problem is also of applied value.
For example, trees might come from an inference method with higher certainty of their branching structure and lower certainty of their nodes order.
A comparison method for such trees should have higher penalty for $\nni$ changes and lower penalty for rank changes, which in our notations requires $\rho < 1$.

In the rest of this section, we show that the $\findpath$ algorithm substantially relies on the fact that the rank move and the $\nni$ move have the same weight in the $\rnni$ graph.
This suggests that a non-trivial algorithmic insight is necessary to extend our polynomial complexity result to other values of $\rho$.

\begin{figure}[ht]
\centering
\includegraphics[width=0.6\textwidth]{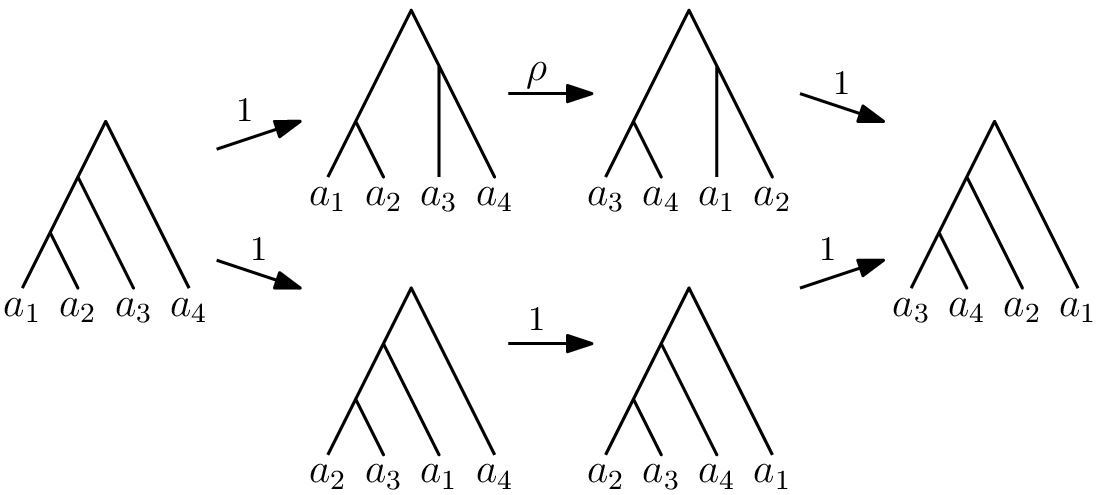}
\caption{Path computed by $\findpath$ (top) and a shorter path (bottom) for $\rho > 1$.}
\label{fig:fp_rho_greater_1_counterexample}
\end{figure}

\begin{proposition}
$\findpath$ does not compute shortest paths in $\rnni(\rho)$ for $\rho \neq 1$.
\label{prop:fp_only_rnni}
\end{proposition}

\proof

For $\rho > 1$ a counterexample is given by the following trees (see Figure~\ref{fig:fp_rho_greater_1_counterexample})
\begin{align*}
	T &= [\{a_1,a_2\},\{a_1,a_2,a_3\},\{a_1,a_2,a_3,a_4\}]\text{ and }\\
	R &= [\{a_3,a_4\},\{a_2,a_3,a_4\},\{a_1,a_2,a_3,a_4\}].
\end{align*}
Applied to these trees $\findpath$ proceeds from $T$ to $[\{a_1,a_2\},\{a_3,a_4\},\{a_1,a_2,a_3,a_4\}]$, then to $[\{a_3,a_4\},\{a_1,a_2\},\{a_1,a_2,a_3,a_4\}]$, and then to $R$.
This path consists of two $\nni$ moves with one rank move in between them and therefore has weight $2 + \rho$.
However, the path from $T$ to $[\{a_2,a_3\},\{a_1,a_2,a_3\},\{a_1,a_2,a_3,a_4\}]$ to $[\{a_2,a_3\},\{a_2,a_3,a_4\},\{a_1,a_2,a_3,a_4\}]$ to $R$ consists of three $\nni$ moves and is hence shorter.

\begin{figure}[ht]
\centering
\includegraphics[width=0.6\textwidth]{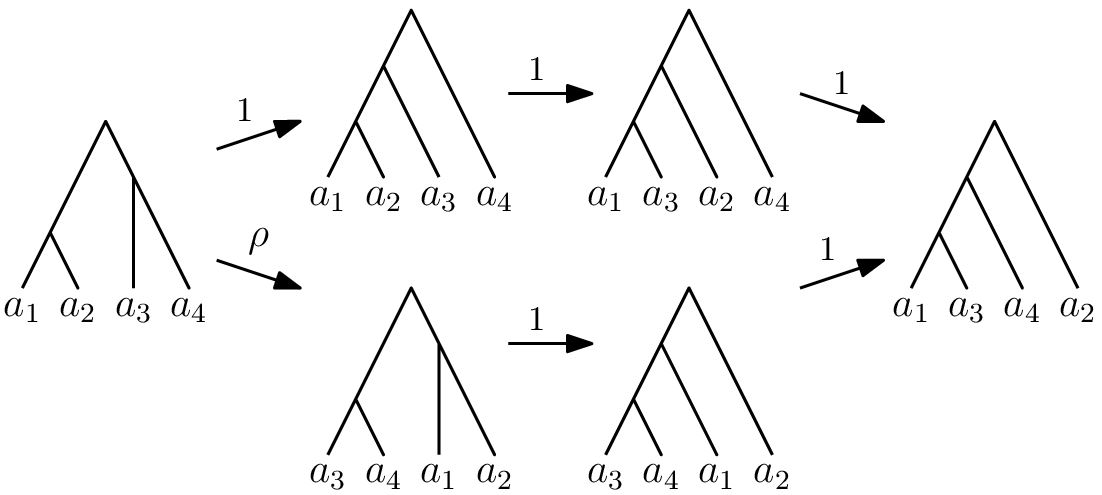}
\caption{Path computed by $\findpath$ (top) and a shorter path (bottom) for $\rho < 1$.}
\label{fig:fp_rho_less_1_counterexample}
\end{figure}

For $\rho < 1$ a counterexample is given by the following trees (see Figure~\ref{fig:fp_rho_less_1_counterexample})
\begin{align*}
	T &= [\{a_1,a_2\},\{a_3,a_4\},\{a_1,a_2,a_3,a_4\}]\text{ and }\\
	R &= [\{a_1,a_3\},\{a_1,a_3,a_4\},\{a_1,a_2,a_3,a_4\}].
\end{align*}
Applied to these trees $\findpath$ proceeds from $T$ to $[\{a_1,a_2\},\{a_1,a_2,a_3\},\{a_1,a_2,a_3,a_4\}]$, then to $[\{a_1,a_3\},\{a_1,a_2,a_3\},\{a_1,a_2,a_3,a_4\}]$, and then to $R$.
This path consists of three $\nni$ moves and therefore has weight $3$.
However, the path from $T$ to $[\{a_3, a_4\},\{a_1, a_2\},\{a_1, a_2, a_3, a_4\}]$ to $[\{a_3,a_4\},\{a_1, a_3, a_4\},\{a_1, a_2, a_3, a_4\}]$ to $R$ consists of one rank move followed by two $\nni$ moves and is hence shorter.
\endproof

\section{Additional open problems}

The idea utilised by \textcite{Dasgupta2000-xa} to prove that computing distances in $\nni$ is $\np$-hard stems from a result that shortest paths in $\nni$ do not preserve clusters \autocite{Li1996-zw}, that is, sometimes a cluster shared by two trees $T$ and $R$ is shared by no other tree on any shortest path between $T$ and $R$.
This counter-intuitive property eventually led to the computational hardness result in $\nni$.
Moreover, this property makes little sense biologically as trees clustering the same set of sequences into a subtree should be closer to each other than to a tree that does not have that subtree.
Indeed, a shared cluster means that both trees support the hypothesis that this cluster has evolved along a subtree.
In light of this biological argument, the $\np$-hardness result can be interpreted as $\decprob{\rho}$ being hard only when the graph $\rnni(\rho)$ is biologically irrelevant.
Although in sharp contrast with the common belief in the field of computational phylogenetics \autocite{Whidden2018-fw}, this interpretation resonates with the idea suggested in the beyond worst case analysis framework \autocite{Roughgarden2019-to} that some problems are only computationally hard when their instances are practically irrelevant.
The following question is hence natural.
\begin{enumerate}
\item For which values of $\rho$ does $\rnni(\rho)$ have the cluster property?
How do those compare to the values of $\rho$ for which $\decprob{\rho}$ is efficient?
\end{enumerate}
Other natural questions that arise in the context of our results are the following.
\begin{enumerate}
\addtocounter{enumi}{1}
\item The questions we have considered for ranked $\nni$ can be studied in other rearrangement-based graphs on leaf-labelled trees, such as the ranked $\spr$ graph and the ranked $\tbr$ graph \autocite{Semple2003-nj}.
What is the complexity of the shortest path problem there?

\item Can our results be used to establish whether the problem of computing geodesics between trees with real-valued node heights is polynomial-time solvable?
This geodesic metric space is called $\mathrm t$-space and an efficient algorithm for computing geodesics in $\mathrm t$-space would be of importance for applications \autocite{Gavryushkin2016-uu}.
\end{enumerate}

\printbibliography
\end{document}